\documentstyle[preprint,aps]{revtex}
\tightenlines
\begin{document}
\title{Effect of higher orbital angular momenta in the baryon spectrum}
\author{H. Garcilazo$^{(1)}$, A. Valcarce$^{(2,3)}$ 
and F. Fern\'andez$^{(2)}$}
\address{$(1)$ Escuela Superior de F\' \i sica y Matem\'aticas \\
Instituto Polit\'ecnico Nacional, Edificio 9,
07738 M\'exico D.F., Mexico}
\address{$(2)$ Grupo de F\' \i sica Nuclear \\
Universidad de Salamanca, E-37008 Salamanca, Spain}
\address{$(3)$ Departamento de F\' \i sica Te\'orica, \\
Universidad de Valencia, E-46100 Valencia, Spain}
\maketitle

\begin{abstract}
We have performed a Faddeev calculation of the baryon spectrum for the chiral
constituent quark model including higher orbital angular momentum states.
We have found that the effect of these states is important, although a
description of the baryon spectrum of the same quality as
the one given by including only the lowest-order 
configurations can be obtained. We have studied
the effect of the pseudoscalar quark-quark interaction on the relative
position of the positive- and negative-parity excitations of the nucleon
as well as the effect of varying the strength of the color-magnetic
interaction.

\vspace*{2cm} \noindent Keywords: \newline
non-relativistic quark models, baryon spectrum, Faddeev calculations \newline
\newline
\noindent Pacs: \newline
12.39.Jh, 14.20.-c
\end{abstract}

\newpage

In a recent paper \cite{GAR1} we calculated the nonstrange baryon spectrum
within a Faddeev approach for the chiral constituent quark model 
\cite{FEVA,VABU,VAGO,FEGO}. This model, besides a confining interaction 
and a perturbative one-gluon exchange, includes a pseudoscalar and a
scalar boson exchange between quarks. In Ref. \cite{GAR1}
only the lowest-order configurations
$(\ell,\lambda,s,t)$ ($\ell$ is the orbital angular momentum
of a pair, $\lambda$ is the orbital angular momentum between the pair and
the third particle, while $s$ and $t$ are the spin and isospin of the 
pair) were included. Thus, for the $N{1/2}^+$ state (corresponding to
the $N$ and to the $N(1440)$) we included the two configurations (0,0,0,0)
and (0,0,1,1). For the $N{1/2}^-$ state (the $N(1535)$) we included
the four configurations (0,1,0,0), (0,1,1,1), (1,0,1,0), and (1,0,0,1), 
while for the $\Delta{3/2}^+$ state (the $\Delta$)
we included the single 
configuration (0,0,1,1). We have now extended our calculation to include all
the configurations $(\ell,\lambda,s,t)$ with $\ell$ and $\lambda$ up
to 5. The objective of the present work is to evaluate the effects of
the higher orbital angular momentum components in a model including 
gluon as well as Goldstone boson exchanges

To study the convergence behavior with respect to the number of
$(\ell,\lambda,s,t)$ configurations we give in Table I the results for
the $\Delta$, $N(1535)$, $N$, and $N(1440)$, corresponding to Fig. 4
of Ref. \cite{GAR1}.
In this table we give the mass difference with respect to the nucleon
ground state including 12 $(\ell,\lambda,s,t)$ configurations, i.e. with 
$\ell$ and $\lambda$ up to 5, considered to be the converged result for the 
nucleon ground state and therefore our mass reference.
As can be seen, the convergence with respect to the
number of configurations is different for different states, so that,
for example, the $\Delta$ comes down by only 5 MeV while the $N(1535)$, 
$N$, and $N(1440)$ they all come down by approximately 100 MeV when one
includes the higher orbital angular momentum configurations. It is also
interesting to notice that while in the lowest-order calculation the
$N(1440)$ lies below the $N(1535)$, when one includes the higher orbital
angular momenta the situation is reversed. We will return to this point 
later on. 
The same study could be done in the case of Figs. 1-3 of Ref. \cite{GAR1}.
However, the results would be similar to those reported on Table I, although
the effect of the higher orbital angular momentum components is smaller in 
these cases.
The reason for this stems
from the fact that the one-pion exchange (OPE) 
force is weaker in these cases, and therefore
the contribution of higher orbital angular momenta is less important.

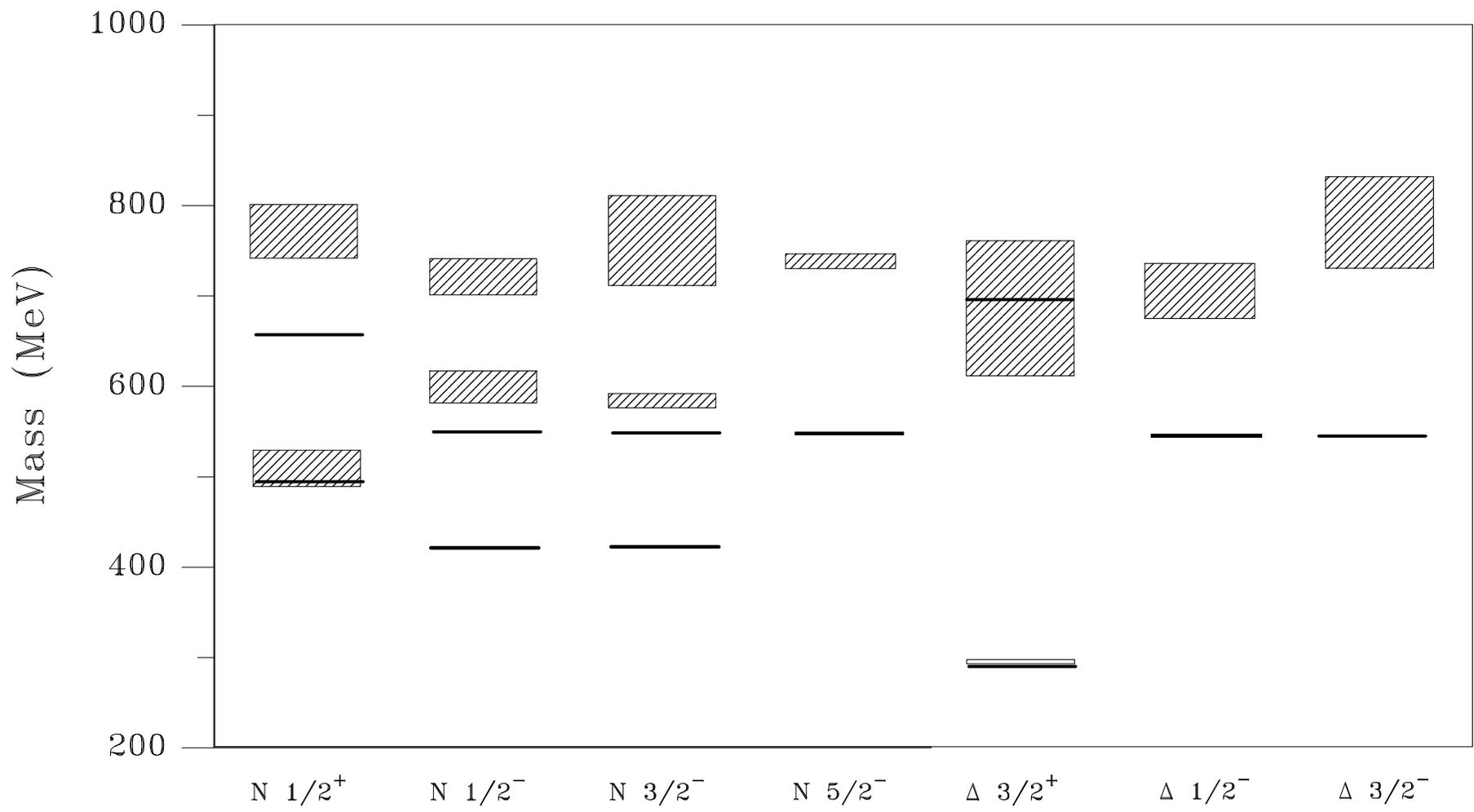
\begin{figure}[tbp]
\caption{ Relative energy nucleon and delta spectrum up to 1.0 GeV
excitation energy for the set of parameters of Table II. The solid
corresponds to the results of the model of Ref. \protect{\cite{GAR1}}
including $\ell$ and $\lambda$ up to 5. The shaded region, whose size
stands for the experimental uncertainty, represents the experimental
data.}
\label{fig1}
\end{figure}

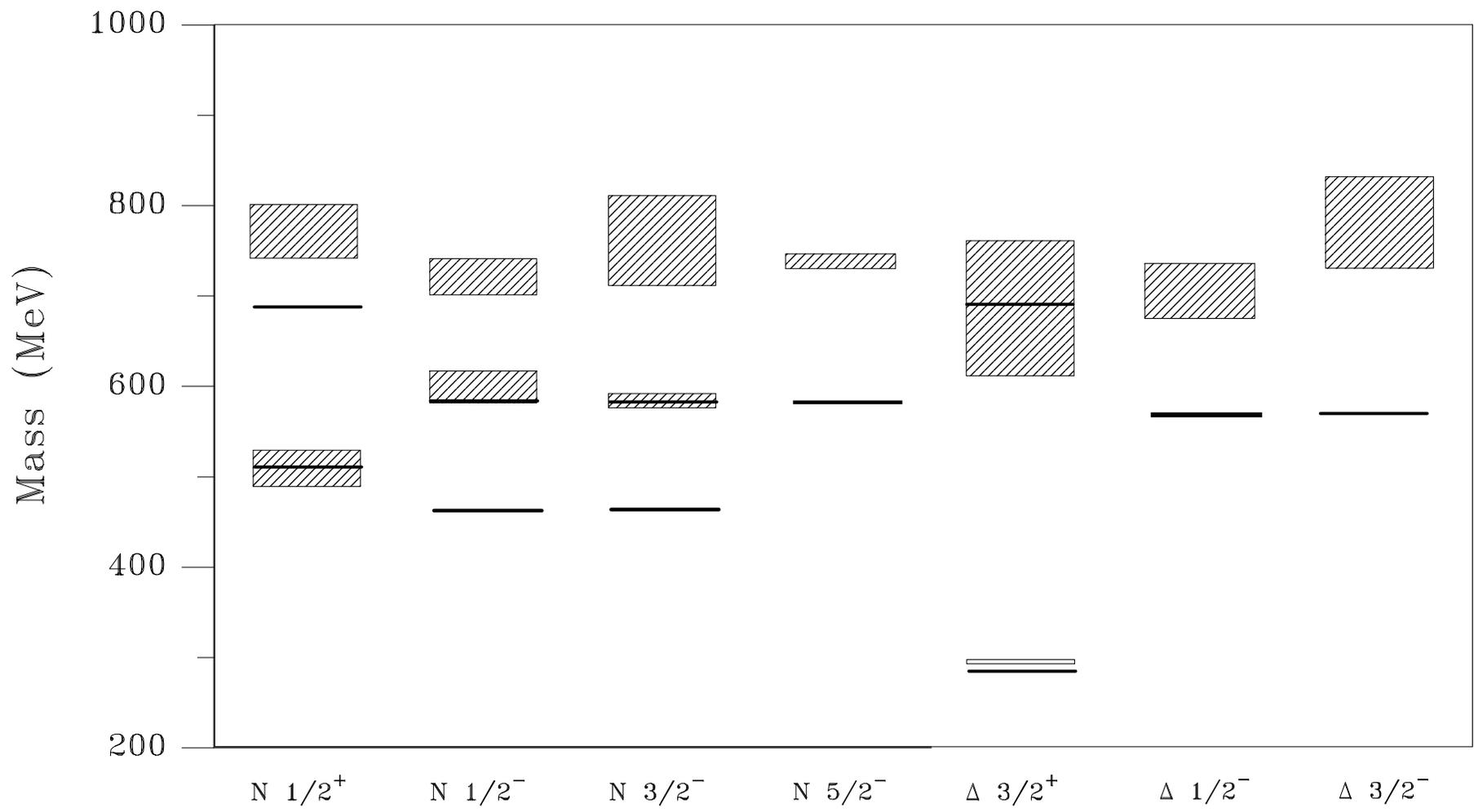
\begin{figure}[tbp]
\caption{ Same as Fig. 1 for the set of parameters of Table III.}
\label{fig2}
\end{figure}

As a consequence of the previous results, it is clear that the
calculation of Ref. \cite{GAR1} was done in a restricted Hilbert
space neglecting orbital angular momentum components that are important
to obtain a conclusion about the performance of the model under discussion.
It therefore arises the question if the model of Ref. \cite{GAR1} can produce
a reasonable baryon spectrum when all the relevant Hilbert space components
are included.
We want to show that similar fits for the baryon spectrum can be
obtained with the extended Faddeev calculation as those obtained with the 
lowest-order version used to calculate the results of Ref. \cite{GAR1}.
As an example, we show in Fig. \ref{fig1} the baryon spectrum obtained using the
parameters of Table II when the higher orbital angular momentum 
configurations are included. 
In this figure we plot by the solid line the excitation energy with 
respect to the nucleon ground state. The shaded areas correspond to the 
experimental value including its uncertainty.
As one can see from this 
figure, the relative position of the positive-parity states $N$, 
$\Delta$, and $N(1440)$ are correctly given, although the negative-parity
states lie below the experimental results as is well-known to happen
for models of this kind. The parameters of Table II, which gave rise to 
this spectrum, are the ones
that have been used in the chiral constituent quark model for the
description of the baryon-baryon interaction \cite{FEVA,VABU}.
The parameter $r_0$ has been taken as 0.25 fm, a typical value
for spectroscopic models, such that the
delta function of the OGE
interaction is regularized in the region of 
short distances. The slope of the confining potential, $a_c$,
which in Table II is 67.0 MeV fm$^{-1}$, is not so different from the 
value 72.5 MeV fm$^{-1}$ used in most of the previous applications
of the model to the baryon spectrum \cite{VAGO,FEGO}.
(One should not forget at this point that consistency with the
two-nucleon sector does not impose any restriction to the value
of $a_c$, because the confining interaction does not contribute
to the nucleon-nucleon potential.) This does not mean,
however, that the fit of the spectrum requires the parameters 
to be close to those of Table II. It is possible to find equally good fits 
with quite different sets of parameters. We show in Fig. \ref{fig2} the
spectrum obtained with the parameters of Table III, which differ 
considerably from those of Table II.

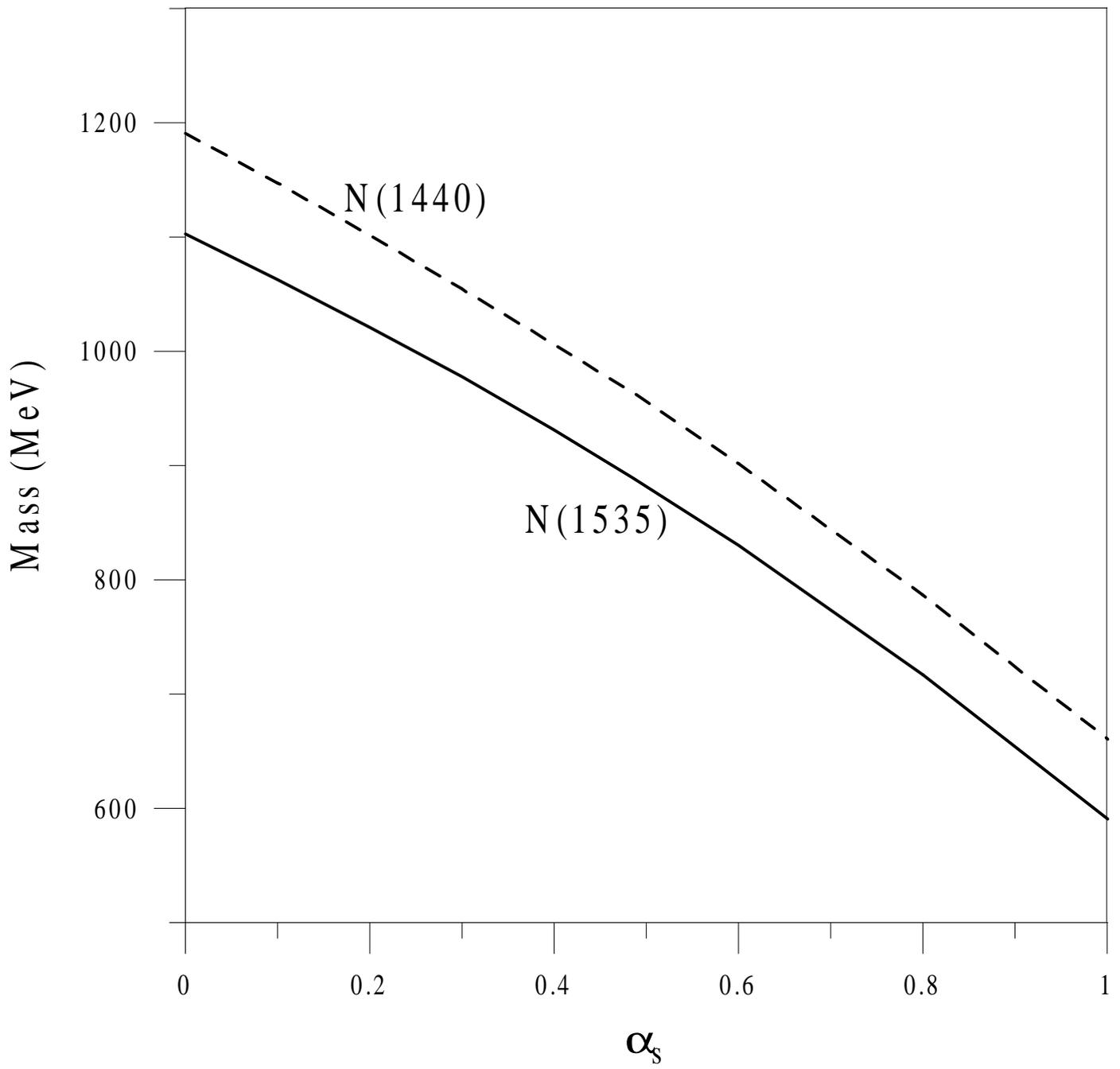
\begin{figure}[tbp]
\caption{$N(1440)$ and $N(1535)$ mass as a function of the
strong coupling constant, $\alpha_s$. See text for details.}
\label{fig3}
\end{figure}

As can be seen in Figs. \ref{fig1} and \ref{fig2}, 
with two different sets of parameters 
for confinement and the one-gluon exchange, fits of a similar quality
are obtained. In both cases one observes how the relative position
of the positive and negative parity excitations of the nucleon are
reversed with respect to the experimental order. This has been stated
as a basic deficiency of models with a strong one-gluon exchange 
interaction. In order to check if the responsible for this incorrect 
ordering is the one-gluon exchange, we have calculated the energy 
of the $N(1440)$ and the $N(1535)$ starting with the set of parameters 
of Table II and varying the contribution of the color-magnetic
interaction by modifying the coupling constant $\alpha_s$ of the
OGE interaction. The results are shown in Fig. \ref{fig3}. As
we are interested in the relative energy of the positive and
negative parity excitations of the nucleon,
we have plotted directly the absolute value obtained from our code. 
In this case the
relative position of the $N(1440)$ and the $N(1535)$ is not very
much affected by the modification of $\alpha_s$. This behavior 
indicates that the correct level ordering between the negative
and positive parity excited states of the nucleon is not  
connected with the lack of OGE.
Moreover, a suppression of the OGE, would imply a stronger pseudoscalar
interaction in order to reproduce the $N-\Delta$ mass difference, and
therefore one would obtain a model that it is incompatible with
the understanding of the basic features of the two-nucleon
system \cite{TOKI}.

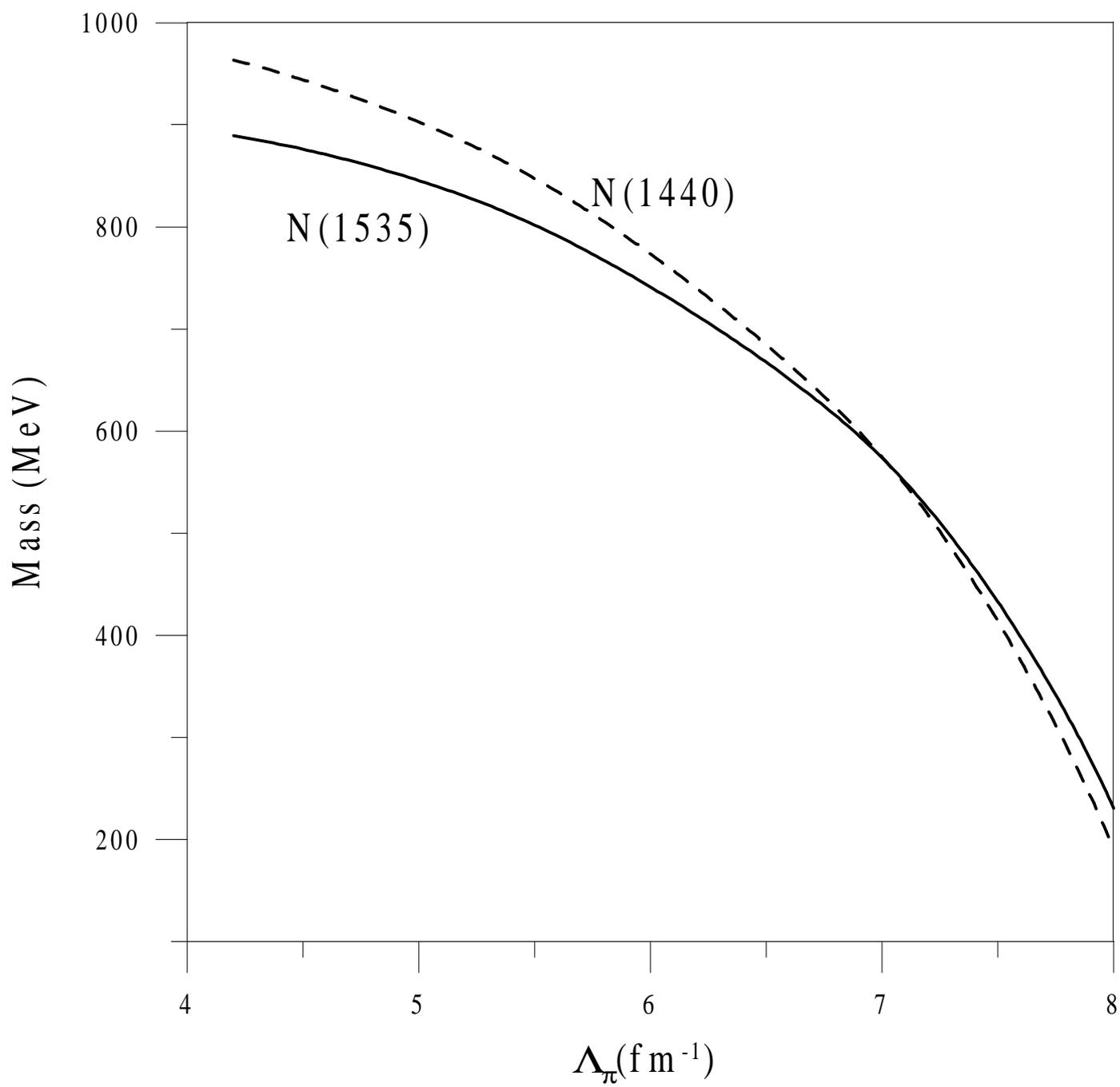
\begin{figure}[tbp]
\caption{$N(1440)$ and $N(1535)$ mass as a function of the
cutoff mass for the one-pion exchange, $\Lambda_\pi$. See text
for details.}
\label{fig4}
\end{figure}

As has been explained in Ref. \cite{GAR1}, a possible mechanism 
responsible for the inversion of these states can be obtained through
the pseudoscalar interaction. To illustrate this point we have 
again calculated the energy of the $N(1440)$ and the $N(1535)$ 
starting with the set of parameters of Table II, but increasing
in this case the contribution of the pseudoscalar interaction by
letting the cutoff parameter $\Lambda_\pi$ of the OPE to increase.
The results are shown in Fig. \ref{fig4}, where we have again
plotted directly the absolute value obtained from our code. 
As can be seen, the inversion of
the ordering between the positive and negative parity states
can be achieved if $\Lambda_\pi$ becomes sufficiently large 
(around 7 fm$^{-1}$ for the set of parameters of Table II). One
should also mention that a model with such a strong cutoff for 
the OPE is not realistic. In this case the $N-\Delta$ mass
difference is around 955 MeV. We have fitted again the
$N-\Delta$ mass difference by modifying the confinement constant
and suppressing the OGE. In this case we lose again the inversion
between the negative and positive parity states. Let at this point
make a comment about the smaller influence of the higher orbital angular
momentum components in the case of Figs. 1-3 of Ref. \cite{GAR1} that we have
mentioned previously. The larger the cut-off of the pseudoscalar interaction, the
larger the nucleon-delta mass difference. This effect is obtained because the 
energy of the nucleon ground state is decreased, enhancing in this way the influence
of the higher orbital angular momentum components.

As a summary, we want to point out that higher orbital angular
momentum components play an important role on the description of the
low-energy baryon spectrum. They mainly influence the relative energy
of the nucleonic and $\Delta$ sectors (as illustrated in Table I), but 
they also affect the relative position of the positive and negative parity
excitations of the nucleon. After reanalyzing the results reported on
Ref. \cite{GAR1}, where these components were not considered, we did not
find any qualitative modification of the conclusions obtained there.
Therefore, the validity of a model where
the OGE is combined with Goldstone-boson exchanges is without
any doubt, and it presents advantages with respect to models 
based only on Goldstone-boson exchanges,
as has been
recently emphasized in several works \cite{TOKI,ISGU}. In particular,
the chiral constituent quark model of Refs. \cite{FEVA,VABU,VAGO,FEGO}
is able to generate a quite reasonable description of the baryon
spectrum, with a set of parameters that allows to understand the
$NN$ phenomenology. For the case of the baryon spectrum,
the set of parameters we found is not unique,
as also stated in Ref.\cite{GAR1}, and we did not find any restriction
to the value of $\alpha_s$. The correct level ordering is closely
related to the pseudoscalar interaction and, in the present model, can
only be achieved at the expense of losing the correct description of the
$NN$ system. As a consequence, other mechanisms should be responsible for
the correct level ordering if consistency with the two-nucleon system 
is required.

\acknowledgements
We like to thank Dr. W. Plessas for enlightening discussions on the 
subject of this paper.
This work has been partially funded by 
COFAA-IPN (Mexico), by Direcci\'{o}n
General de Investigaci\'{o}n Cient\'{\i}fica y T\'{e}cnica (Spain) 
under the Contract No. PB97-1410, by
Junta de Castilla y Le\'{o}n under the Contract No. SA109/01. 
A. V. thanks the Ministerio de Educaci\'on, 
Cultura y Deporte for finantial support through the Salvador 
de Madariaga program.

\begin{table}
\caption{Convergence of baryon masses with respect to the number (Nr.)
of configurations $(\ell,\lambda,s,t)$. See text for details.}    
\label{table1}
\begin{tabular}{cccccccc}
 Nr. & $M_\Delta$(MeV) & Nr. & $M_{N(1535)}$(MeV) &  Nr. & $M_N$(MeV)  
& Nr.  & $ M_{N(1440)}$(MeV) \\
\tableline
 1 & 423 & 4  & 643  & 2   & 115 & 2  & 641 \\
 2 & 419 & 8  & 601  & 4   & 68  & 4  & 622 \\
 3 & 418 & 12 & 570  & 6   & 21  & 6  & 587 \\
   &     & 16 & 551  & 8   & 12  & 8  & 578 \\
   &     & 20 & 540  & 10  & 2   & 10 & 562 \\
   &     &    &      & 12  & 0   & 12 & 558 \\
\end{tabular}
\end{table}

\begin{table}
\caption{ Quark model parameters for the calculation of Fig. 1.}
\label{table2}

\begin{tabular}{cccc}
 & $m_q ( {\rm MeV})$                          &  313       & \\
 & $\alpha_s$                                  &  0.485      & \\
 & $a_c ( {\rm MeV} \cdot {\rm fm}^{-1})$      &  67.0     & \\
 & $\alpha_{ch}$                               &  0.0269    & \\
 & $r_0 ({\rm fm})$                            &  0.25       & \\
 & $m_\sigma ({\rm fm}^{-1})$                  &  3.42      & \\
 & $m_\pi ({\rm fm}^{-1})$                     &  0.7       & \\
 & $\Lambda_\pi ({\rm fm}^{-1})$               &  4.2       & \\
 & $\Lambda_\sigma ({\rm fm}^{-1})$            &  4.2       & \\
\end{tabular}
\end{table}

\begin{table}
\caption{ Quark model parameters for the calculation of Fig. 2.}
\label{table3}

\begin{tabular}{cccc}
 & $m_q ( {\rm MeV})$                          &  313       & \\
 & $\alpha_s$                                  &  0.75      & \\
 & $a_c ( {\rm MeV} \cdot {\rm fm}^{-1})$      &  60.12     & \\
 & $\alpha_{ch}$                               &  0.0269    & \\
 & $r_0 ({\rm fm})$                            &  0.8       & \\
 & $m_\sigma ({\rm fm}^{-1})$                  &  3.42      & \\
 & $m_\pi ({\rm fm}^{-1})$                     &  0.7       & \\
 & $\Lambda_\pi ({\rm fm}^{-1})$               &  4.2       & \\
 & $\Lambda_\sigma ({\rm fm}^{-1})$            &  4.2       & \\
\end{tabular}
\end{table}

\end{document}